# Landau-Zener-Stückelberg interferometry in multilevel superconducting flux qubit

Xueda Wen,[1] Yiwen Wang,[1] Shanhua Cong,[2] Guozhu Sun,[2] Jian Chen,[2]
Lin Kang,[2] Weiwei Xu,[2] Yang Yu,[1, *] Peiheng Wu,[2] and Siyuan Han[2, 3]

[1]*National Laboratory of Solid State Microstructures and Department of Physics, Nanjing University, Nanjing 210093, China*
[2]*Research Institute of Superconductor Electronics and Department of Electronic Science and Engineering, Nanjing University, Nanjing 210093, People's Republic of China*
[3]*Department of Physics and Astronomy, University of Kansas, Lawrence, KS 66045, USA*

Landau-Zener-Stückelberg interferometry has been extensively investigated in quantum two-level systems, with particular interests on artificial system such as superconducting flux qubits. With increasing the driving field amplitude, more energy levels will be involved into the quantum evolution, which results in population inversion and many interesting interference patterns. These interference patterns can be used to obtain the parameters characterizing the system and probe dephasing mechanisms of the qubit. Most recently, experiments have been extended to the regime with higher-frequency and larger-amplitude driving field, in which the interference pattern exhibited more complicated characteristics. In this article, we give a universal description of the characteristics observed in both low-frequency and high-frequency regimes. Besides explaining the already observed experimental results, our theoretical model predicted many interesting phenomenon, which can be demonstrated by future experiments.

## I. Introduction

Superconducting Josephson devices recently emerged as a platform for exploring coherent quantum dynamics in solid state systems [1–3]. Compared with natural atoms and molecules, these devices can be strongly coupled to external radio-frequency (rf) fields while preserving quantum coherence [4–6]. A large number of experiments associated with strong driving, e. g., Rabi oscillations in the multiphoton regime [7–9], Bloch oscillations [10], and dressed states of superconducting qubits under extreme driving [11], have been done on these devices. Recently, coherent dynamics of superconducting qubits in the regime dominated by Landau-Zener (LZ) transitions were extensively studied [12–23]. One may use LZ transitions to enhance the quantum tunneling rate [19, 20], prepare the quantum state [21], control the qubit gate operations effectively [22], and so on. Moreover, consecutive LZ transitions can induce Landau-Zener-Stückelberg (LZS) interference [24]. Since the theoretical scheme of observing LZS interference in qubits was proposed by Shytov *et al.* [16], a series of experiments on LZS interference were implemented in two-level systems (TLS) such as flux qubits [13, 15, 18, 23] and charge qubits [14], which provided an alternative method to manipulate and characterize qubit in the strongly driven regime.

A new regime of strong driving was developed in recent works [25, 26]. Unlike previous experiments which employed one TLS [13–15], the qubit in this experiment was driven through a manifold of several states spanning a wide energy range. The population of the qubit under large-amplitude fields exhibited a series of diamond like interference patterns in the space parameterized by flux detuning and microwave amplitude. The interference patterns, which displayed a multi-scale character, encoded the information of several energy levels of the system. Most recently, experiments were extended to the regime with higher-frequency driving field [27, 28]. The experimental results revealed many interesting characteristics due to high frequency driving field and the participation of higher energy levels. In this article, we give a universal description of LZS interference phenomenon that have been observed in superconducting flux qubit in both low- and high-frequency regimes. We emphasize that although we focus on superconducting flux qubit here, the theoretical model can be applied to other systems that have similar energy structures. This article is organized as follows. In Sec. II, we give a general description of the dynamics in multilevel superconducting flux qubit driven by large amplitude field. In Sec. III, we apply the theoretical model to the regime with low-frequency driving field. We show that the theoretical results are agreed with experiments perfectly. In Sec. IV, we use the same model to explain the most recent experiments which employed large-frequency driving field. In Sec. V, based on our theoretical model, we predict more interesting phenomenon that can be demonstrated by future experiments. In Sec. VI, we show that our theoretical model can be generalized to a variety of quantum systems.

## II. General Model and Method

---

*Electronic address: `yuyang@nju.edu.cn`

In this section, starting from the simple TLS model, we first obtain the LZS induced transition rate between two quantum states that share an avoided level crossing formed by the coupling between two states. Then we construct the dynamical process with rate equations, which is appropriate in dealing with stationary population distribution.

A driven two-level system subject to the effects of decoherence can be described by the Hamiltonian [12, 15, 29]:

$$\hat{H}(t) = -\frac{\Delta}{2}\hat{\sigma}_x - \frac{h(t)}{2}\hat{\sigma}_z, \quad (1)$$

where $\Delta$ is the tunnel splitting; $\hat{\sigma}_x$ and $\hat{\sigma}_z$ are Pauli matrices. $h(t)$ is the time dependent energy detuning from an avoided crossing:

$$h(t) = \epsilon + A\sin\omega t + \delta\epsilon_{noise}(t), \quad (2)$$

where $\epsilon$ is the dc component of the energy detuning, $A$ and $\omega$ are the amplitude and frequency of the driving rf field respectively, $\delta\epsilon_{noise}(t)$ is the classical noise. As discussed in ref [14], by using white noise model and perturbation theory, one can obtain the LZS transition rate between the states $|0\rangle$ and $|1\rangle$:

$$W(\epsilon, A) = \frac{\Delta^2}{2}\sum_n \frac{\Gamma_2 J_n^2(x)}{(\epsilon - n\omega)^2 + \Gamma_2^2} \quad (3)$$

where $\Gamma_2 = 1/T_2$ is the dephasing rate and $J_n(x)$ are Bessel functions of the first kind with the argument $x = A/\omega$. Eq.(3) implies that the transition rate is proportional to $\Delta^2$ which is decided by the energy structure of the system. Extending Eq.(3) to multilevel superconducting flux qubit [25], we can write the LZS transition rate between diabatic quantum states $|i, L\rangle$ (left well, $i = 0, 1, 2...$) and $|j, R\rangle$ (right well, $j = 0, 1, 2...$) as:

$$W_{ij}(\epsilon_{ij}, A) = \frac{\Delta_{ij}^2}{2}\sum_n \frac{\Gamma_2 J_n^2(x)}{(\epsilon_{ij} - n\omega)^2 + \Gamma_2^2}, \quad (4)$$

where $\Delta_{ij}$ is the avoided crossing between states $|i\rangle$ and $|j\rangle$, and $\epsilon_{ij}$ is the dc energy detuning from the corresponding avoided crossing $\Delta_{ij}$. The time evolution of population for state $|i, L\rangle$ can be described by a rate equation:

$$\begin{aligned}\dot{P}_{i,L} =& -\sum_j W_{ij} P_{i,L} + \sum_j W_{ij} P_{j,R} \\ & -\sum_{i'} \Gamma_{i\to i'} P_{i,L} + \sum_{i'} \Gamma_{i'\to i} P_{i',L} \\ & -\sum_j \Gamma_{i\to j} P_{i,L} + \sum_j \Gamma_{j\to i} P_{j,R},\end{aligned} \quad (5)$$

where $\Gamma_{i\to i'}$ is the intrawell relaxation rate from $|i, L\rangle$ to $|i', L\rangle$, and $\Gamma_{i\to j}$ is the interwell relaxation rate from $|i, L\rangle$ to $|j, R\rangle$. The time evolution of population for state $|j, R\rangle$ can be obtained in the same way. In the stationary case, we have $\dot{P}_{i,L(R)} = 0$. The qubit population distribution in each well can be easily obtained from

$$P_{L(R)} = \sum_{i(j)} P_{i,L(j,R)} \quad (6)$$

### III. Applications in Low Frequency Regime

In this section, we apply the model and method presented in Sec.I to the regime where the flux qubit is driven by low-frequency and large-amplitude field. As observed in experiments [25, 26], the qubit population distribution reveals a series of diamond-like patterns. In this section, we first give a clear picture of the underlying physics that governs the interference patterns, and then make a quantitative study of the experimental results. For clear, we analyze the first two 'diamonds' separately.

#### (a) First Diamond

Shown in FIG.1(a) is a schematic energy diagram of a multilevel superconducting flux qubit. The initial state is prepared at ground state $|0R\rangle$. By applying microwave, the system is driven through avoided level-crossing $\Delta_{0R,0L}$ repetitively, LZS interference will occur between states $|0R\rangle$ and $|0L\rangle$. This is the familiar LZS interference in TLS, as shown in region I in FIG.1(c). With further increase of the amplitude of driving field to reach the avoided level-crossing $\Delta_{1R,0L}$ (see also the dotted line in FIG.1(c)), LZS interference between states $|1R\rangle$ and $|0L\rangle$ will occur. In this case, the population for state $|0L\rangle$ may be pumped to $|1R\rangle$, and subject a rapid relaxation to the ground state $|0R\rangle$, leading to a diminution of population in left well, as shown in region II in FIG.1(c).

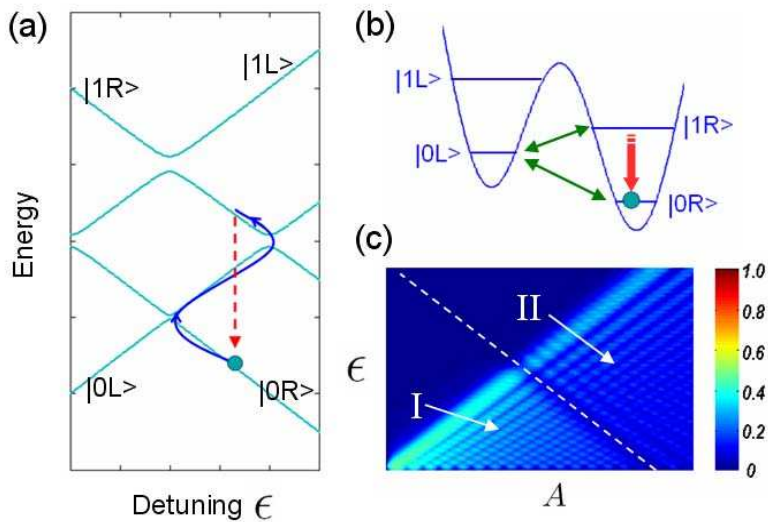

FIG. 1: (a) Schematic energy diagram and different transition processes of a superconducting flux qubit. (b) Schematic drawing of different transition processes in double well configuration. (c) Calculated qubit population versus energy detuning and driving amplitude. The dotted line indicates where the field amplitude reaches $\Delta_{0L,1R}$. On the right of the dotted line, there is an obvious population diminution due to the transitions $|0L\rangle \to |1R\rangle \to |0R\rangle$.

The physics picture discussed above can be described by a group of rate equations:

$$\begin{cases} \dot{P}_{0R} = -W_{0R,0L}P_{0R} + \Gamma_{1R\to 0R}P_{1R} + (W_{0R,0L} + \Gamma_{0L\to 0R})P_{0L} \\ \dot{P}_{1R} = -(W_{0L,1R} + \Gamma_{1R\to 0R})P_{1R} + W_{0L,1R}P_{0L} \\ P_{0R} + P_{0L} + P_{1R} = 1. \end{cases} \quad (7)$$

In the stationary case, Eqs.(7) can be easily solved. The calculated results with experimental parameters are shown in FIG.1(c), which agree with experiments very well [25, 30].

### (b) Second Diamond

There are two main processes generating the second diamond:

(i) When the amplitude of driving field is further increased to reach the avoid level crossing $\Delta_{0R,1L}$, LZS interference between states $|0R\rangle$ and $|1L\rangle$ can occur. Since the LZS transition rate is proportional to $\Delta_{i,j}^2$ (see Eq.(4)), and $\Delta_{0R,1L}^2 \gg \Delta_{0R,0L}^2$ in superconducting flux qubit system, the dynamical process in this region is dominated by the transitions between states $|0R\rangle$ and $|1L\rangle$ (see FIG.2(a)). Then the population inversion observed in this region is easy to understand. At the begining the population at $|0R\rangle$ is pumped to $|1L\rangle$, and then decays to $|0L\rangle$ rapidly. Because $|0L\rangle$ is metastable ($\Gamma_{0L\to 0R} \ll W_{0R,1L}, \Gamma_{1L\to 0L}$), the population is concentrated at $|0L\rangle$ and population inversion appears.

(ii) For the population concentrated at $|0L\rangle$, LZS interference at avoided level crossing $\Delta_{0L,1R}$ can occur. For $\Delta_{0L,1R} = \Delta_{0R,1L}$, this transition process cannot be neglected. In short, the population at $|0L\rangle$ is pumped to $|1R\rangle$, and then decays rapidly to $|0R\rangle$ (see FIG.2(a)(b)). This process breaks the population inversion induced by process (i).

It is the interplay between processes (i) and (ii) that leads to the population inversion and checkerboard pattern in the second diamond. To do a quantitative study, we employ rate equations to describe the processes discussed above:

$$\begin{cases} \dot{P}_{0R} = -W_{0R,1L}P_{0R} + \Gamma_{1R\to 0R}P_{1R} + \Gamma_{0L\to 0R}P_{0L} + W_{0R,1L}P_{1L} \\ \dot{P}_{0L} = -(W_{0L,1R} + \Gamma_{0L\to 0R})P_{0L} + W_{0L,1R}P_{1R} + \Gamma_{1L\to 0L}P_{1L} \\ \dot{P}_{1R} = -(W_{0L,1R} + \Gamma_{1R\to 0R})P_{1R} + W_{0L,1R}P_{0L} \\ P_{0R} + P_{1R} + P_{0L} + P_{1L} = 1. \end{cases} \quad (8)$$

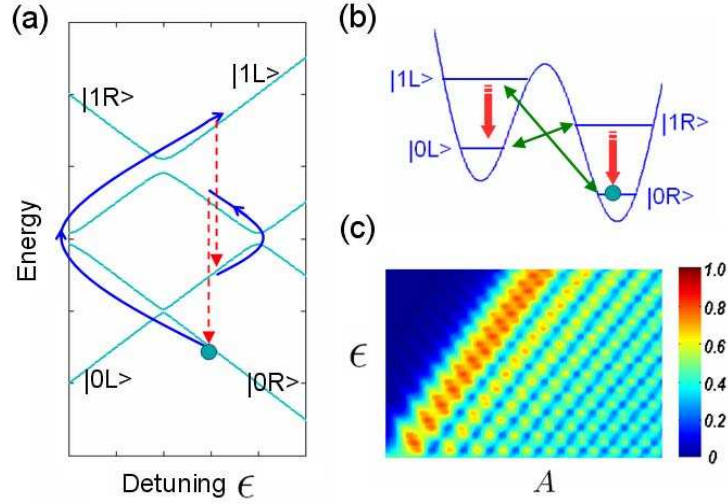

FIG. 2: (a) Schematic energy diagram and different transition processes of a superconducting flux qubit. (b) Schematic drawing of different transition processes in double well configuration. (c) Calculated qubit population versus energy detuning and driving amplitude. The population inversion is induced by the process $|0R\rangle \to |1L\rangle \to |0L\rangle$, and broken by the process $|0L\rangle \to |1R\rangle \to |0R\rangle$.

In the stationary case $\dot{P} = 0$, Eqs.(8) can be easily solved. The theoretical results (see FIG.2(c)) which show both population inversion and checkerboard pattern are perfectly agreed with experiments.

### (c) Multi − Diamond

With further increase the amplitude of driving field, more energy levels will participate in the quantum evolution. As observed in experiments [25], the qubit population distribution reveals a series of diamonds. The diamond structures result from the interplay between energy detuning and driving amplitude, which mark when the various level crossings are reached. Because the onset of each diamond is associated with transitions at a particular level crossing, the boundaries of the diamonds mark the occurrence of level crossings. Theoretically, these patterns can be obtained by extending Eqs.(5) to more energy levels. The underlying physics is similar with those in (a) and (b). Shown in FIG.3 is the calculated result with Eqs.(5) by considering ten energy levels. The main features agree with experiments very well [25].

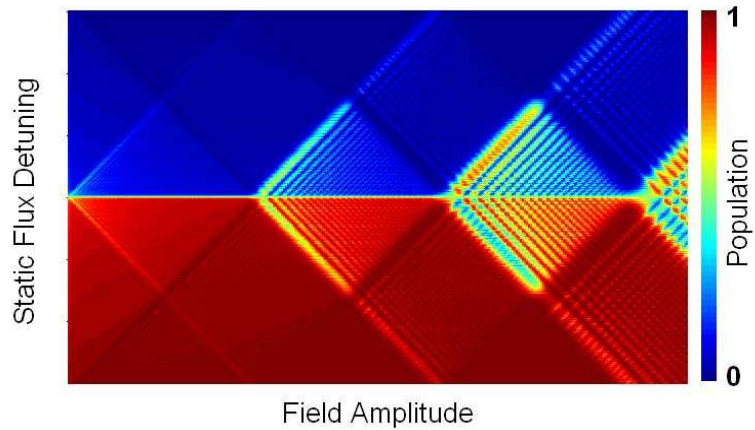

FIG. 3: Calculated qubit population versus energy detuning and driving amplitude. The simulation parameters are from experiments.[25]

### IV. Applications in High Frequency Regime

Before we discuss the difference of interference patterns in low frequency regime and high frequency regime, we first define two characteristic physical quantities:

(i) Span of resonance peak $\delta A$. For $n-$photon resonance, the transition rate depends on field-amplitude $A$ in the form

$$W_n(A) \propto J_n^2(A/\omega). \tag{9}$$

The span of resonant peak can be evaluated from the zeros of $J_n(x)$ (for example, the zeros of $J_0(x)$ are $x \simeq 2.40, 5.52, \ldots$). Then it is easy to find that the characteristic span of resonance peak has the simple form $\delta A \sim \omega$.

(ii) Distance of sequential diamonds $\delta D$. From the analysis in Sec. II, we know that the boundaries of the diamonds mark the occurrence of level crossings. Therefore, the distance of sequential diamonds equals the distance of two sequential avoided level crossings, i.e., $\delta D \sim |D_{0R,nL} - D_{0R,(n+1)L}|$, where $D_{0R,nL}$ and $D_{0R,(n+1)L}$ are the locations of $\Delta_{0R,nL}$ and $\Delta_{0R,(n+1)L}$, respectively.

From the analysis above, it is easy to find that when $\delta A \sim \delta D$, i.e., $\omega \sim \delta D$, sequential diamonds will merge together. We define the frequency in this regime as *high frequency*. We emphasize that comparing with the large energy separation far from the avoided level crossing, the high frequency is actually *not* high because the LZS interference picture still holds in this regime. Recently, experiments in this regime have been reported in two works (see FIG.4(b) and FIG.5(b)) [27, 28]. Especially, as shown in FIG.4(b), it is obvious that two sets of interference patterns (marked by red arrows and black arrows) merge together, as we have discussed above. Theoretically, the experimental results can be easily obtained by extending our rate equation method to high frequency regime. As shown in FIG.4(c) and FIG.5(c), the positions of interference patterns obtained from our rate equation method agree with experimental results.

Comparing the two experiments in high frequency regime [27, 28], it is interesting to find that the experimental results are sensitive to the barrier height. Generally, the barrier height affects experimental results mainly in two aspects: (i) A higher barrier allows more local quantum states participating in LZS interference, which results in rich interference patterns. On the contrary, if the barrier is too low, only a few local quantum states exist. Driven by large amplitude field, the population will mostly be pumped to non-local states, and relax to both wells with equal probability, resulting in no observable effects in experiments. This is why in experiments [25, 27, 28] the interference patterns become no longer distinguishable when the field amplitude becomes larger. (ii) A high barrier means a suppressed quantum tunneling effect, i.e., a small coupling $\Delta_{ij}$ between $|iL\rangle$ and $|jR\rangle$. In this case, the interference effect will be suppressed, and only LZS interference at larger $\Delta_{ij}$ will be easily observed. Therefore, under the interplay between (i) and (ii), there is only a certain window for observing LZS interference patterns. This is why in FIG.4(b) two sets of interference patterns were observed while in FIG.5(b) only one set of interference patterns were observed.

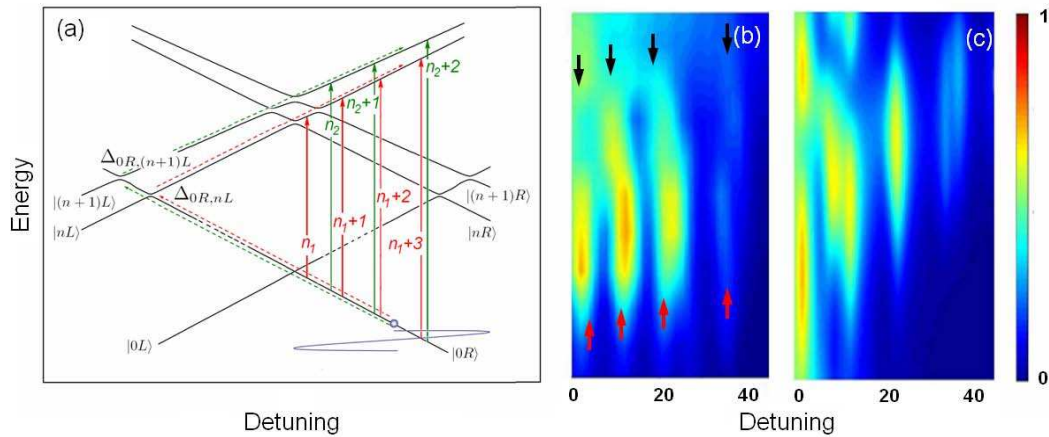

FIG. 4: (a) Energy-level diagram illustrating the LZS interference processes. Driven by harmonic external fields, LZS interference occurs at the avoided level crossings $\Delta_{0R,nL}$ and $\Delta_{0R,(n+1)L}$. (b) Measured dependence of qubit population in the left well on energy detuning and field amplitude. (c) Calculated qubit population distribution(for details of experimental and simulation parameters, please refer to[28]). LZS interference at $\Delta_{0R,iL}(i > n+1)$ leads to a population transition to non-local quantum state, resulting in indistinguishable patterns, corresponding to the larger amplitude region in (b). LZS interference at $\Delta_{0R,iL}(i < n)$ makes little contribution to interference patterns because of small $\Delta_{0R,0L}$.

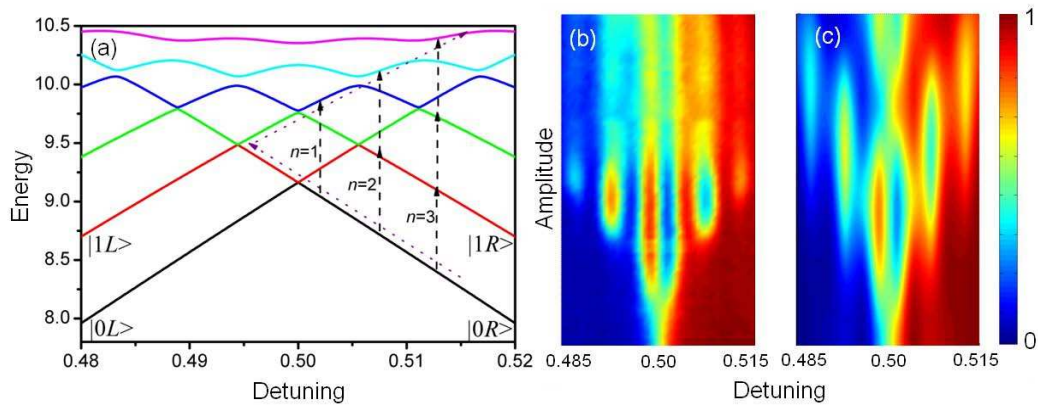

FIG. 5: (a) Energy-level diagram illustrating the LZS interference processes. Driven by harmonic external fields, LZS interference occurs at the avoided level crossings $\Delta_{0R,1L}$. (b) Measured dependence of qubit population in the left well on energy detuning and field amplitude. (c) Calculated qubit population distribution(for details of experimental and simulation parameters, please refer to[27]). LZS interference at $\Delta_{0R,iL}(i > 2)$ leads to a population transition to non-local quantum state, resulting in indistinguishable patterns, corresponding to the larger amplitude region in (b). LZS interference at $\Delta_{0R,iL}(i = 1)$ makes little contribution to interference patterns because of small $\Delta_{0R,iL}$.

## V. More Interesting Interference Patterns

Assured by the excellent agreement with experiments, we can utilize our theoretical model to predict more interesting interference patterns, which is of great meaning in manipulating qubit population. Shown in FIG.6 are examples of simulated interference patterns based on experimental parameters [28], some of them have been experimentally demonstrated [31]. We emphasize that decoherence induced by environments also contributes a lot to the the richness of interference patterns, which can be discussed by using the same theoretical model [30].

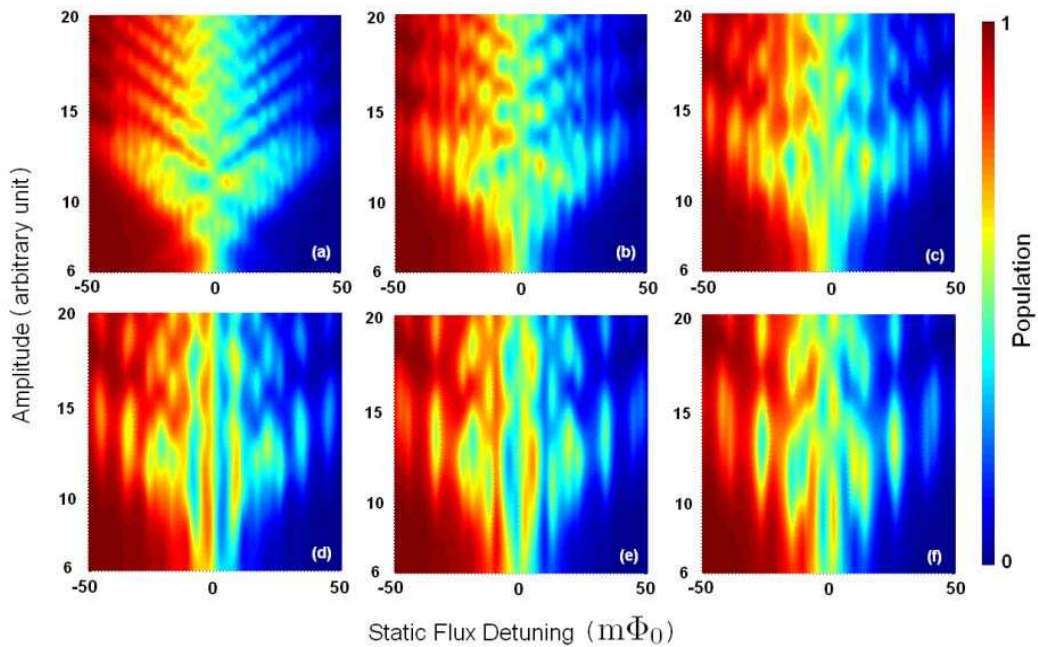

FIG. 6: Calculated qubit population versus energy detuning and driving amplitude. The simulation parameters are from experiments[28]. The driving frequencies are (a)5GHz, (b)8GHz, (c)11GHz, (d)13GHz, (e)15GHz, and (f)17GHz, respectively.

## VI. Conclusion

We have utilized rate equation method to investigate the LZS interference observed in superconducting flux qubits in both low frequency and high frequency regime, and our theoretical results agree with experiments very well. Based on our theoretical model, we can predict more interesting interference patterns, which is important for characterizing the parameters and decoherence of qubits. The LZS interferometry also provide an alternative method to manipulate qubit population in experiments. In addition, our theoretical model can be generalized to other quantum systems with similar energy structures such as superconducting charge qubit [14] and superconducting qubit and microscopic two-level system coupling system. Especially, the later system has received dramatic attention recently because it is proposed that microscopic two-level system themselves can serve as qubits while the superconducting qubit can perform as data bus [32–35]. The coupled qubit microscopic two-level system has the very similar energy structure as flux qubits, and we believe that our theoretical model will be of great use in investigating the quantum information process in such systems.